\input phyzzx
\overfullrule=0pt
\nopagenumbers
\line{\hfil BROWN-HET-1106}
\line{\hfil December 1997}
\vskip .50in
\title{{\bf LIGHT-FRONT PARTONS AND DIMENSIONAL\hfill
 REDUCTION IN RELATIVISTIC FIELD THEORY}
\foot{Work supported in part by the Department
of Energy under contract DE-FG02-91ER40688 - Task A}}
\vskip .75in
\author{{\bf Antal JEVICKI}}
\centerline{{\it Department of Physics}}
\centerline{{\it Brown University, Providence, RI  02912, USA}}
\vskip 1.0in
\abstract
\singlespace
\noindent  We present a simple discussion of the appearance 
of light-front partons  in local field theory.The
description in terms of partons provides a dimensional reduction 
which relates a 2+1 with a 3+1 dimensional theory for example. The
possibility for existence of Lorentz symmetry and a connection
to the relativistic membrane is described.It is shown how the
reconstruction of the full relativistic field theory is possible
with a proper treatment of the parton configuration space. Compared with
the case of identical particles this involves keeping 
configurations where the partons are at the same points.

\endpage

\pagenumbers
\normalspace

{\bf 1.}  It has been originally suggested by t'Hooft$^1$ that in order to reconcile
gravity with quantum mechanics there should be a reduction of degrees of
freedom from 3+1 to 2+1 dimensions.  This reduction, operating as a 
holographic mapping should allow for an inverse namely a recovery of
 the 3d data from 2d.
t'Hooft has presented a model based on cellular automata as an
implementation of this idea. One of the main challenges is the demonstration of 
Lorentz invariance in 3+1D. In several papers Susskind has emphasized$^2$ the use of 
light-front quantization and partons for 
obtaining the holographic projection.  Indeed a string  theory in the light
cone frame can be formulated$^{3,4}$ without the explicit use of a longitudinal dimension. In
general one has the basic idea of light-front partons which would
serve as most
fundamental degrees of freedom. This would among others significantly improve the ultraviolet properties of the
theory and allow a consistent microscopic understanding of the physics
of black holes.The recently introduced M(atrix) models belong to the above
category of theories$^{5,6}$. 

In this note we point out that a similar mechanism could be operational already in local field theory. We present a  simple discussion of the 
appearance of light-front partons and the corresponding dimensional reduction  
in the example of  a  relativistic scalar field. Our discussion will be at classical level with the interactions ignored. This will be sufficient for
our purpose of presenting the basic ideas . We are then able to describe a mechanism for recovering the full theory from the lower dimensional (parton) 
picture. 

The initial idea is to consider a relativistic field in the light-front 
frame and
after compactifying the longitudinal direction $x^-$
truncate the fields by keeping only oscillators with
the lowest value of $p^+$ .  These would
represent the basic partonic degrees of freedom and the question is if 
it is possible to build the full theory in terms of them.  We first show that even though this represents a seemingly huge truncation, one still has the possibility
of a 3+1 Lorentz symmetry in the emerging 2+1 dimensional theory. We give
a set of generators closing a 4D Poincare algebra. These are closely
related to generators of the $4D$ relativistic membrane and we explain in some detail this 
connection.  
 We then  proceed to our main topics and
study the $N$-body problem of free partons.  We show that using appropriately
defined (collective) fields the full $4D$ relativistic scalar field theory is reconstructed.This reconstruction is based on a special treatment of the
$N$-parton configuration space.As opposed to the standard case of
identical particles we include in this space configurations of
coinciding partons. This comes in analogy with a similar treatment of 
light cone strings. The fact that a relativistic field can be represented in terms of
nonrelativistic partons and described in lower dimensional terms
is likely to give us new insight into the properties of local field theory.
\vskip .10in
{\bf 2.}  Consider a relativistic scalar field theory $\varphi (x_{\mu} )$ in 3+1
dimensions
$$
{\cal L} = {1\over 2} \, \partial_{\mu} \varphi \, \partial^{\mu} \varphi - V
(\phi )
$$
On the light-front plane $x^+ = {\rm const.}\,\, (x^{\pm} = {1\over\sqrt{2}}
\left( x^0 \pm x^3 )\right)$ one has the Dirac brackets
$$\left\{ \varphi ( x^- , x ) , \varphi (y^- , y) \right\} = \left(\partial_-
\right)^{-1} \delta (x^- - y^- ) \cdot \delta (x-y)
$$
Here and in what follows $x = (x^1 , x^2 ) $ stands for all the transverse directions.  The Hamiltonian
and the longitudinal momentum (generating translation in $x^+$ and $x^-$
respectively) read
$$
\eqalign{ H \equiv P^- & = \int dx^- d^2 x \left\{ {1\over 2} \nabla \varphi \nabla
\varphi + V (\varphi ) \right\} \cr
P^+ & = \int dx^-d^2 x \partial_- \varphi \partial_- \varphi }
\eqno\eq
$$
When we compactify the $x^-$ direction for convenience with a radius $R = 2\pi
$, the conjugate momenta $p^+ = {2\pi n\over R}$ are given by  positive integers.
Expanding in the $p^+$ modes:
$$
\varphi (x^- , x) = \sum_{n>0}  {1\over\sqrt{4\pi p_n^+}} \, \left( A_n (x) e^{-i
p_{n}^{+} x^{-}} + A_n^+ (x) e^{ip_{n}^{+} x^{-}} \right) + {1\over
\sqrt{2\pi}} \tilde{\phi} (0,x) \eqno\eq
$$
In order to obtain something like the parton picture with fundamental degrees of freedom we now
truncate the system to just the $p^+ = 1$ mode (we ignore all the other modes
including the zero mode):
$$
\varphi (x^- , x) \approx {1\over\sqrt{4\pi}} \left( A_1 (x) e^{-i x^{-}} + A_1^+
(x) e^{ix^{-}}\right)
$$
The Hamiltonian and the longitudinal momentum then reduce to
$$
\eqalign{ H & = \int d^2 x \, \left( {1\over 2}\nabla A^+ (x) \nabla A + V\right) \cr
 P^+ & = \int d^2 x \, A^+ (x) A(x) }
\eqno\eq$$
where we have dropped the index $(p^+ = 1 )$ from our creation-annihilation
operators. In general it is expected that if one interprets the reduction as integrating 
out degrees of freedom that one would have changes to the interaction
term.
The creation-annihilation operators obey the Poisson brackets of
a nonrelativistic second quantized Schrodinger field
$$
\left\{ A (x) , A^+ (x') \right\} = \delta^{(2)} (x - x')\eqno\eq
$$
and give the density function  $\rho (x)
\equiv A^+ A$.
We see that the longitudinal momentum is the integral of the density and equal to the number operator
$$ P^+ = \int d^2 x  \rho (x) = \hat{N}
$$
Changing to the density and phase variables
$$\eqalign{ A (x) & = \sqrt{\rho} \, e^{i\pi}\cr
A^+ (x) & = \sqrt{\rho} \, e^{-i\pi} }
$$
we write the Hamiltonian as
$$
H = \int d^2 x \left({1\over 2}  \rho \nabla \pi \nabla \pi + V' (\rho ) \right)
$$
where an additional $\rho$-dependent term was added to the potential energy.  As
we have mentioned, we will ignore interactions and $\hbar$-quantum effects in
what follows.

To conclude, by truncating the relativistic field $\varphi (x^- , x)$ to just
the $p^+ = 1$ mode we have a nonrelativistic type theory in 2+1 dimensions with
$\rho (x)$ and $\pi (x)$ as conjugate fields. These are the density fields
of partons.

We will now point out that there can exist a 4D Lorentz symmetry still 
operating in this reduced theory.  It is given by the following set of generators
$$\eqalign{ J^{ab} & = \int d^2 x \, \pi \left( x^a \partial_b - x^b \partial_a
\right)\, \rho \qquad a,b = 1,2\cr
J^{a-} & = \int d^2 x {1\over 2} \left\{ x^a \rho \left( \nabla \pi \right)^2 - \rho
\partial_a (\pi^2 ) \right\}\cr
J^{a+} & = \int d^2 x x^a\rho\cr
J^{+ -} & = -\int d^2 x\, \rho \pi }
$$
These together with
$$\eqalign{ P^- &  = \int d^2 x {1\over 2} \, \rho \left( \nabla \pi \right)^2\cr
P^+ & = \int d^2 x \, \rho\cr
P^a & = \int d^2 x \pi \partial_a \, \rho }
$$
can be seen to close the 3+1 dimensional Poincare algebra.

This algebra is directly related to the Poincare algebra of the 4D membrane
$(X^{\mu} (\tau, \sigma_1 , \sigma_2 ) ; \mu = 0 , 1, 2, 3 )$ given by
Bordemann and Hoppe in$^7$.  In the light cone gauge after fixing the area
preserving diffeomorphisms, there remains a scalar field $(q (x) , p (x))$
and also the zero modes $(x_0^- , p_0^+ )$ corresponding to the $X^- (\sigma)$
and $P^+ (\sigma)$ coordinates
$$\eqalign{
&\left\{ q (x) , p (x') \right\} = \delta^{(2)} (x - x' )\cr
& \left\{ x_0^- , p_0^+ \right\} = -1 }
$$
The full set of Poincare generators  constructed in$^7$ is in terms of these
sets of variables.While our fields $(\rho (x) , \pi (x))$ are analogous to
the continuous variables$(q (x) , p(x))$ of the membrane we do not have any
discrete zero modes $(x_0^- , p_0^+ )$.It is somewhat puzzling
that a Lorentz symmetry still exists even without the longitudinal
zero modes. This can be explained as follows:

The Hamiltonian in the case of the membrane reads
$$
P^- = {1\over 2p_0^+} \, \int d^2 x \, q \left( (\nabla \pi )^2 + {1\over q^2}
\right)
$$
The other generators and especially the most relevant longitudinal boost
operator $J^{a-}$  involve
in a nontrivial way both the discrete $(x_0^- , p_0^+ ) $ and the continuum
$(q
, p)$ degrees of freedom.  One notices however a scaling a symmetry in the membrane Hamiltonian:
$$\eqalign{
q & \rightarrow \lambda q \qquad p \rightarrow \lambda^{-1} p\cr
x_0^- & \rightarrow \lambda x_0^- \qquad p_0^+ \rightarrow \lambda^{-1} p_0^+ }
$$
generated by
$$
Q_s = p_0^+ \, x_0^- - \int q p d^2 x
$$
Using this symmetry, one can perform a Hamiltonian reduction\foot{This
observation is due to Jean Avan.} setting
$$
Q_s = 0 \quad {\rm and} \quad p_0^+ = 1
$$
This results in elimination of the zero mode degrees of freedom leaving just
the 2d field $(q (x) , p(x))$ which is identified with our $(\rho (x) , \pi (x)
)$.  Incidentally this reduction also  explains  the sometimes
mysterious identification
$$
P^+ = \int d^2 x \rho = N
$$
made in $M$(atrix) theory .
\vskip .10in

{\bf 3.}  We would now like to demonstrate that it is possible to recover the full
(relativistic) field theory from the $N$-body system of partons.  Consider the
kinetic energy
$$
H = \sum_{i=1}^N \, {1\over 2} \, p_i^2
$$
where the coordinates are in $d = D-2$ dimensions
$$
x_i = \left( x_i^1 , x_i^2 , \cdots x_i^d \right)
$$
The time conjugate to the Hamiltonian is the null-front time $x^+ =  x^0
+ x^{d+1}$ and consequently $H = P^-$.  The total longitudinal momentum is
$$
P^+ = \sum_{i=1}^N 1 = N
$$

We will now define a sequence of collective fields $\rho_n (x)$ on the
coordinate space
$$
X^d \times X^d \times \cdots \times X^d / S_N
$$
by treating this space as an orbifold\foot{After completion of this work we have
learned of a similar orbifold treatment of the $N=1$ instanton moduli space by J
Brodie and S. Ramgoolam$^{10}$} in parallel with the case of strings and the
string light cone $N$-body dynamics as given in [8,9].

The idea is to treat in a special way the
configurations when locations of partons coincide.  It is well known
that in the standard treatment of identical particles$^{10}$ one excludes
from the connfiguration space the points
when the particles are at the same location.  In the present orbifold logic,
these points are included and reinterpreted as different observables.
This then allows us to construct a sequence of (collective) density fields as follows.  First
$$
\rho_1 (x) = \sum_i ^{\prime} \, \delta (x - x_i )
$$
with the prime denoting the fact that in the sum we only have {\bf non coincidental} 
coordinates.  To define this we can simply introduce an infinitesimal  distance $\epsilon$ and
define a minimal separation by $\vert x_{i'} - x_{i''} \vert > \epsilon$.  In turn, if
two coordinates are within $\epsilon$ we will treat this as a composite object
with $p^+ = 2$.  Consequently all pairs of coordinates such that
$$
\vert x_{i'} - x_{i''} \vert < \epsilon
$$
will be taken to contribute to
$$\rho_2 (x) = \sum_{(i', i'')} \, {1\over 2} \left( \delta (x - x{_i'} ) +
\delta (x - x_{i''} )\right)
$$
Generally
$$
\rho_n (x) = \sum_{(i_1 , i_2, \cdots i_n )} {1\over n} \left( \sum_{k=1}^n \,
\delta (x - x_{i_{k}} )\right)
$$
which receives a contribution only if some $n$- of the particles are at the
same point.  This defines the $p^+ = n$ density field.  The sequence of
densities so constructed is to be directly identified with the
creation-annihilation
operators in the expansion of the relativistic field $\varphi (x^-  ,x)$.More precisely 
$$\rho_n (x) = A_n^+ (x) A_n (x)
$$
and we have correctly reconstructed the total longitudinal momentum operator
$$
P^+ = \sum_{n>0} n \int \rho_n (x) d x = \sum_{n>0} n\int A_n^+ A_n dx^d
$$
What is less trivial is the Hamiltonian.  Here we use the methods of [12] and
write
$$
\sum_{i=1}^N \, {\partial^2\over\partial x_i^2} = \sum_n \sum_i \int  dx dy
{\partial\rho_n (x)\over \partial x_i} \, {\partial \rho_n (y)\over \partial
x_i} \, {\partial\over \partial \rho_n (x)} \, {\partial\over\partial\rho_n
(y)} + {\partial^2 \rho_n\over \partial x_i^2} \, {\partial\over\partial
\rho_n}
$$
The second term is known to lead to potential type contributions so let us just
follow the first kinetic term (for details see$^{12}$).A simple calculation
gives
$$
\sum_i p_i^2 \rightarrow \int d^2 x \sum_n \, {1\over n} \, \rho_n (x)
\nabla \pi_n (x)  \nabla \pi_n (x)  + \cdots
$$
so that with
$$
A_n = \sqrt{\rho_n} \, e^{i\pi_{n}} , A_n^+ = \sqrt{\rho_n} e^{-i\pi_{n}}
$$
we have
$$
P^- =\sum  {1\over 2} p_i^2 \rightarrow \int d^d x \sum_n {1\over 2n} \, \nabla
A_n^+ \nabla A_n = \int dx^- dx^d \, \nabla\varphi\nabla\varphi
$$
We have shown than that the relativistic field theory expression is recovered from the
nonrelativistic parton picture.
 
To summarize the basic point of this treatment rests on the question of statistics
obeyed by the so called parton coordinates. For a  usual system of identical 
particles we exclude from the coordinate space configurations where
the particles coincide, indeed this assumption is crucial for the existence of 
anyon statistics in 2d for example. If for the light front partons we
decide to keep the configurations with coinciding points these configurations then
carry additional information: that is the origin for a sequence of
density fields of growing $p^+$ which we have built to reconstruct the
single relativistivistic field. If on the other hand partons obey some 
standard statistics this construction would not be applicable.

The present picture of local fields is now in line with that of (light-cone)strings.
One
hopes that this can play a role in reconciling the loop properties of
point particles and strings.One of the basic properties of closed string loops
is modular invariance,a symmetry which has no counterpart in field theory.
We have presented our construction with little attention to interactions
and one would clearly like to have a full interacting theory at hand. The 
Lorentz generators of the membrane allow an additional singular interaction,
this would represent a fine tunning or possibly a critical point in the
relativistic scalar case. A much more interesting example which we would
like to bring up is the theory of self-dual gravity which is known 
to posses certain large N composite structure$^{13}$.  
\vskip .10in

\noindent{\bf References}
\medskip

\pointbegin
G. t Hooft,  ``Dimensional Reduction in Quantum Gravity", Essay dedicated to Abdus Salam, (gr-gc/9310006).
\point
L. Susskind,  ``The World as a Hologram", {\it J. Math. Phys.} {\bf 36} 11 (1995).  
\point
C. Thorn,  ``In Sakharov Conference on Physics", Moscow, Vol 81, (1991).
\point
I. Klebanov and L. Susskind, {\it Nucl. Phys.} {\bf B309}, 175 (1988).
\point
T. Banks, W. Fischler, S. Skenker and L. Susskind,hep-th/9610043.
\point
L. Susskind, (hep-th/9704080).
\point
M. Bordemann and J. Hoppe, {\it Phys. Lett.} {\bf B} 317, 315 (1993)
\point
R. Dijkgraaf, E. Verlinde and H. Verlinde, {\it Nucl. Phys.} {\bf B500}, 
43-61 (1997) hep-th/9702187;R.Dijkgraaf,G.Moore,E.Verlinde and H.Verlinde,
hep-th/9608096.
\point
L. Motl,"Proposal on Nonperturbative Superstring Interactions" hep-th/9701025;
T.Banks and N.Seiberg"Strings from Matrices",hep-th/9702187.
\point
J. Brodie and S. Ramgoolam,hep-th/9711001. 
\point J. M. Leinaas and J. Myrheim,  ``On the Theory of Identical Particles", {\it Il Nuovo Cimento}, Vol 37, No.1, 132 (1977).
\point
A. Jevicki and B. Sakita, {\it Nucl. Phys.} {\bf B165}, 511 (1980) .
\point 
A. Jevicki, M. Mihailescu and J. Nunes, hep-th/9706223 ,work in progress.

\bye